\documentclass{aastex}
\usepackage{emulateapj5,apjfonts,epsfig}

\slugcomment{Accepted for publication in ApJ Letters}

\shorttitle{OPTICAL COUNTERPART OF SAX J1808.4--3658}
\shortauthors{WANG ET AL.}

\begin{document}

\title{The Optical Counterpart of the Accreting Millisecond Pulsar 
SAX~J1808.4--3658 in Outburst: Constraints on the Binary Inclination}

\author{Zhongxiang Wang,\altaffilmark{1}
   Deepto~Chakrabarty,\altaffilmark{1}
   Paul~Roche,\altaffilmark{2} 
   Philip~A.~Charles,\altaffilmark{3} 
   Erik~Kuulkers,\altaffilmark{4}
   Tariq~Shahbaz,\altaffilmark{5}
   Chris~Simpson,\altaffilmark{6}
   Duncan~A.~Forbes,\altaffilmark{7}
   and Stephen~F.~Helsdon\altaffilmark{8}
}

\altaffiltext{1}{Department of Physics and Center for
	Space Research, Massachusetts Institute of Technology, Cambridge,
	MA 02139; wangzx@space.mit.edu, deepto@space.mit.edu} 

\altaffiltext{2}{Department of Physics and Astronomy, University of Leicester,
        Leicester LE1 7RH, UK; pdr@star.le.ac.uk}

\altaffiltext{3}{Department of Physics and Astronomy, University of
        Southampton, Southampton SO17 1BJ, UK; pac@astro.soton.ac.uk}

\altaffiltext{4}{Space Research Organization of the Netherlands
	(SRON), Sorbonnelaan 2, 3584 CA Utrecht, The Netherlands;
	E.Kuulkers@sron.nl} 

\altaffiltext{5}{Current address: Instituto de Astrofisica de
        Canarias, E38200 La Laguna, Tenerife, Canary Islands, Spain;
        tsh@ll.iac.es} 

\altaffiltext{6}{Subaru Telescope, National Astronomical Observatory
        of Japan, 650 North A'ohoku Place, Hilo, HI 96720; chris@naoj.org}

\altaffiltext{7}{Centre for Astrophysics and Supercomputing, Swinburne
        University of Technology, Hawthorn, VIC 3122, Australia;
        dforbes@swin.edu.au} 

\altaffiltext{8}{School of Physics and Astronomy, University of Birmingham,
        Edgbaston, Birmingham B15 2TT, UK; sfh@star.sr.bham.ac.uk}

\begin{abstract}
We present multiband optical/infrared photometry of V4580 Sgr, the
optical counterpart of the accretion-powered millisecond pulsar
SAX~J1808.4--3658, taken during the 1998 X-ray outburst of the system.
The optical flux is consistent with emission from an X-ray--heated
accretion disk.  Self-consistent modeling of the X-ray and optical
emission during the outburst yields best-fit extinction
$A_V=0.68^{+0.37}_{-0.15}$ and inclination $\cos
i=0.65^{+0.23}_{-0.33}$ (90\% confidence), assuming a distance of
2.5~kpc.  This inclination range requires that the pulsar's stellar
companion has extremely low mass, $M_{\rm c}=$0.05--0.10~$M_\odot$.
Some of the infrared observations are not consistent with disk
emission and are too bright to be from either the disk or the
companion, even in the presence of X-ray heating.
\end{abstract}


\section{INTRODUCTION}

It is generally believed that millisecond radio pulsars are formed
during sustained mass transfer onto neutron stars in X-ray binaries
(e.g., Bhattacharya \& van den Heuvel 1991).  Only one example of a presumed
progenitor, an accretion-powered millisecond X-ray pulsar, is
currently known.  The X-ray transient SAX~J1808.4--3658
($l=355.4^\circ$, $b=-8.1^\circ$) was discovered
in 1996 September by the {\em BeppoSAX} Wide Field Cameras during a
$\sim$20 day transient outburst (in't Zand et al. 1998).  Based on the
detection of Eddington-limited thermonuclear X-ray bursts during these
observations, the source distance is estimated to be 2.5~kpc (in~'t
Zand et al. 2001).  A second source outburst was detected with the
{\em Rossi X-ray Timing Explorer (RXTE)} in 1998 April (Marshall
1998).  Timing analysis of the 2--30 keV {\em RXTE} data revealed the
presence of a 401~Hz pulsar in a 2-hr binary with a low-mass companion
(Wijnands \& van der Klis 1998; Chakrabarty \& Morgan 1998).

Shortly after the initial {\em RXTE} detection of the 1998 X-ray outburst,
we observed a $V\approx16$ star located 19 arcsec from the center of
the {\em BeppoSAX} error circle; this star was not present on the
Digitized Sky Survey image of the field to a limiting magnitude of
$V\gtrsim 19$, leading to its identification as the optical
counterpart of SAX J1808.4--3658 (Roche et al. 1998).  The optical
intensity of this source faded as the X-ray source declined, and a
2-hr orbital modulation was marginally detected in the optical flux
(Giles, Hill, \& Greenhill 1999).   This 2-hr optical modulation was
subsequently confirmed in observations during quiescence (Homer et
al. 2001).  The optical counterpart has been designated V4580
Sagittarii (Kazarovets, Samus, \& Durlevich 2000).  In this {\em Letter}, 
we report on optical/IR photometry obtained during the 1998 outburst.

\section{OBSERVATIONS AND RESULTS}

We obtained multiband optical photometry of the~SAX J1808.4--3658
field at several epochs during the 1998 X-ray outburst using the
$f/15$ Cassegrain CCD imager on the 1-m Jacobus Kapteyn Telescope
(JKT) at the Observatorio del Roque de los Muchachos, La Palma, Canary
Islands, Spain.  
Additional optical observations were obtained at various epochs during
the outburst using the Keck 10-m telescope in Mauna Kea, Hawaii; the
3.5-m New Technology Telescope (NTT) at the European Southern
Observatory (ESO) in La Silla, Chile; and the 1.9-m telescope at the
South African Astronomical Observatory.  Infrared photometry was also
obtained at several epochs using the 3.8-m United Kingdom Infrared
Telescope (UKIRT).  A summary of these observations is given in
Table~1.  For completeness, we have also included the photometry
obtained by other groups as well (Giles et al. 1999; Percival et
al. 1998; Homer et al. 2001).
\begin{figure*}[t]
\centerline{\epsfig{file=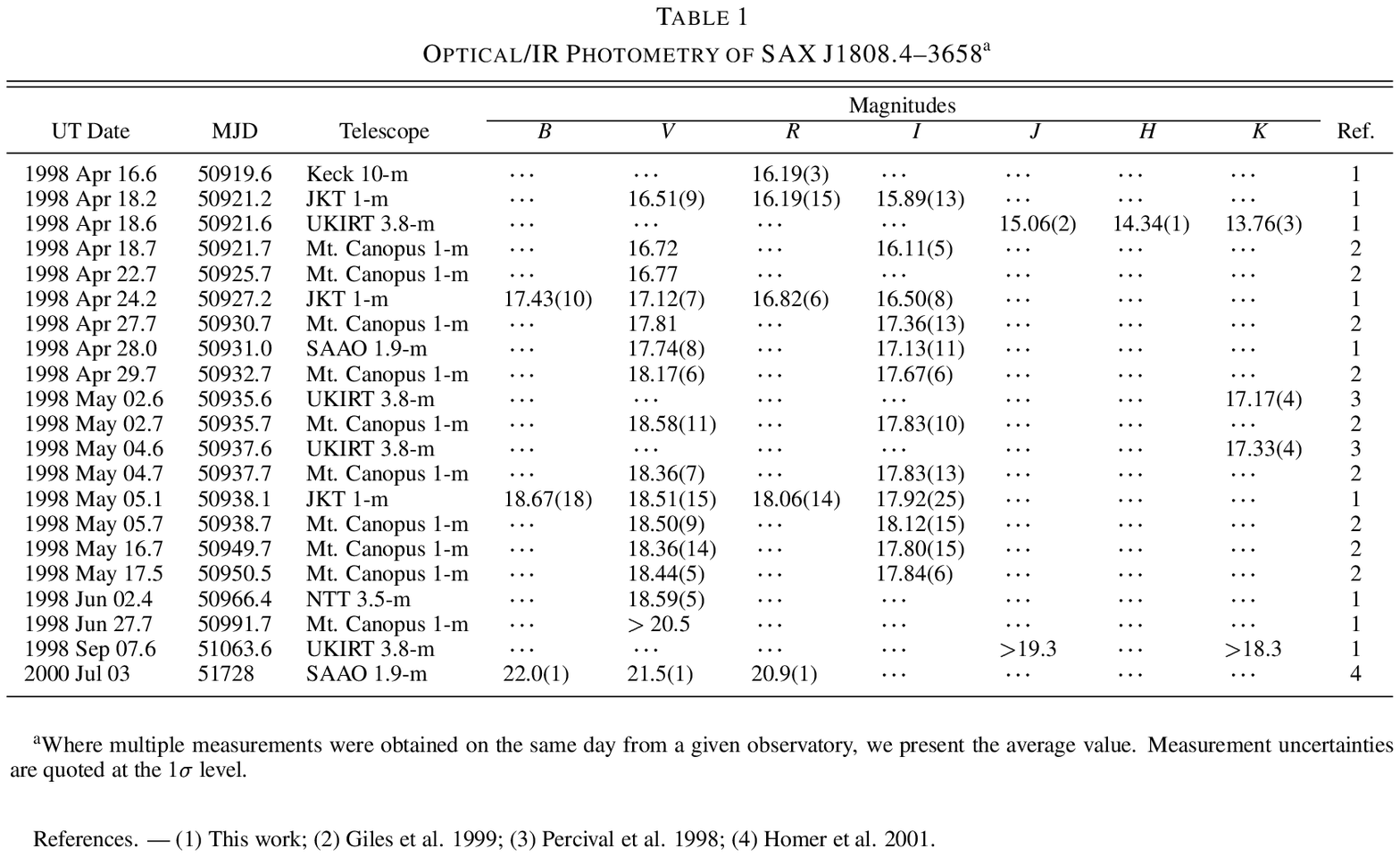}}
\vspace*{0.12in}
\end{figure*}

The $V$-band flux history is shown in Figure~1, along with the X-ray
flux history measured by the {\em RXTE} Proportional Counter Array
(PCA; Gilfanov et al. 1998).  On both plots, we have also indicated
the quiescent flux levels measured well after outburst (Wijnands et al.
al. 2001; Homer et al. 2001).   It is interesting to compare the
behavior of the X-ray and optical light curves.  In the X-ray band,
the intensity shows a steady exponential decay ($\tau=10.9$ d) until
about MJD 50929, when there is a sharp break to a steeper decay
($\tau= 2.2$ d), as shown previously by Gilfanov et al. (1998).
The optical $V$-band light curve also shows an initially exponential
decay in intensity ($\tau=8.4$ d) until MJD 50936, when it abruptly
reaches a plateau lasting at least 30~d.  Despite the fact that the
optical light curve appears to roughly follow the X-ray light curve
early in the outburst, the breaks from the initial decay are spaced by
a week, and the behavior after the break is quite different in the two
bands.    

The broadband optical/IR spectrum is shown in Figure 2 for several
epochs during the outburst, as well as a quiescent measurement.  The
shape of the {\em BVRI} spectrum during the outburst remains roughly
constant, and we show below that this shape is easily consistent with
an X-ray heated disk model.  There is an obvious infrared excess on
MJD 50921, with the {\em JHK} points lying well above an extrapolation
of the {\em BVRI} spectrum.  By contrast, the $K$ point for MJD 50938
is consistent with the extrapolated optical spectrum.  Since the
origin of the IR excess early in the outburst is unclear, we restrict
our accretion disk fitting in the next section to the optical {\em
BVRI} bands, which (along with the ultraviolet and soft X-ray) is
where most of the disk emission from an LMXB accretion disk is
expected.

One point of concern is the discrepancy between optical intensities
measured at different observatories a short time apart.  This is
particularly evident in the JKT and Mt. Canopus observations on MJD
50921, which were spaced by half a day but differ by 0.2 magnitudes in
$V$ and $I$.   One possible contribution to this discrepancy is the
2-hr orbital flux modulation reported by Giles et al. (1999), with
amplitude of $\simeq 0.07$ magnitudes.  To account for this in our
fits, we added a systematic uncertainty of this size in quadrature to
the statistical uncertainties quoted in Table~1.  However, even this
systematic uncertainty is insufficient to explain the discrepancy,
which we conclude largely arises from further systematic errors due to
calibration uncertainties in the JKT data.  In particular, several of
the JKT service observations suffered from limited or insufficient
calibration measurements.   We therefore adopt a systematic
uncertainty of 0.2 magnitudes for all the JKT measurements in our
model fitting.

\section{SPECTRAL FITTING}

The optical data during outburst are well fitted by an X-ray--heated
accretion disk model (see Vrtilek et al. 1990, Chakrabarty 1998, and
references therein).  We summarize this model here.  The observed flux
from the accretion disk at frequency $\nu$ can be written as
\begin{equation}
  F_\nu = \frac{4\pi\,h\,\nu^3\,e^{-A_\nu/1.086}\,\cos i}{c^2 D^2}
    \int_{r_{\rm in}}^{r_{\rm out}}\frac{r\,dr}{\exp [h\nu/kT(r)] - 1} ,
\end{equation}
where $D$ is the source distance, $A_\nu$ is the
(frequency-dependent) interstellar extinction in magnitudes, $r$ is
the mid-plane disk radius coordinate, $r_{\rm in}$ and $r_{\rm out}$
are the inner and outer radii of the disk, and the disk's surface temperature
profile $T(r)$ is given by
\begin{equation}
   T^4(r) = \frac{3 G M_{\rm x}\dot M}{8\pi\sigma r^3}
          + \frac{L_{\rm x}(1-\eta_{\rm d})}{4\pi\sigma r^2}
            \left(\frac{dH}{dr} - \frac{H}{r}\right) .
\end{equation}
Here, $M_{\rm x}$ is the neutron star mass, $\dot M$ is the mass
transfer rate through the disk, $L_{\rm x}$ is the X-ray luminosity of
the neutron star, $\eta_{\rm d}$ is the X-ray albedo of the disk, and $H$ is
the disk's scale height.  $H(r)$ may be determined from the condition of
hydrostatic equilibrium, as in a standard Shakura-Sunyaev disk model
(see, e.g., Frank, King, \& Raine 1992).  The first term in equation
(2) is due to internal viscous heating in the disk, and the second
term is due to X-ray heating.  

For SAX~J1808.4--3658, many of these model parameters are
well-constrained.  We may adopt the distance $D=2.5$ kpc inferred from
radius-expansion X-ray bursts (in~'t Zand et al. 2001).  Since this is
a disk-accreting pulsar, we may assume that the inner disk is
truncated by the pulsar's magnetosphere at a radius $r_{\rm in}$ of
order the corotation radius $r_{\rm co}=(GM_{\rm x}P_{\rm
spin}^2/4\pi^2)^{1/3} \approx 30$ km (Psaltis \& Chakrabarty 1999); in
fact, the optical spectrum is not very sensitive to the exact value of
this parameter, since the optical emission primarily arises from radii
in excess 

\centerline{\epsfxsize=8.5cm\epsfbox{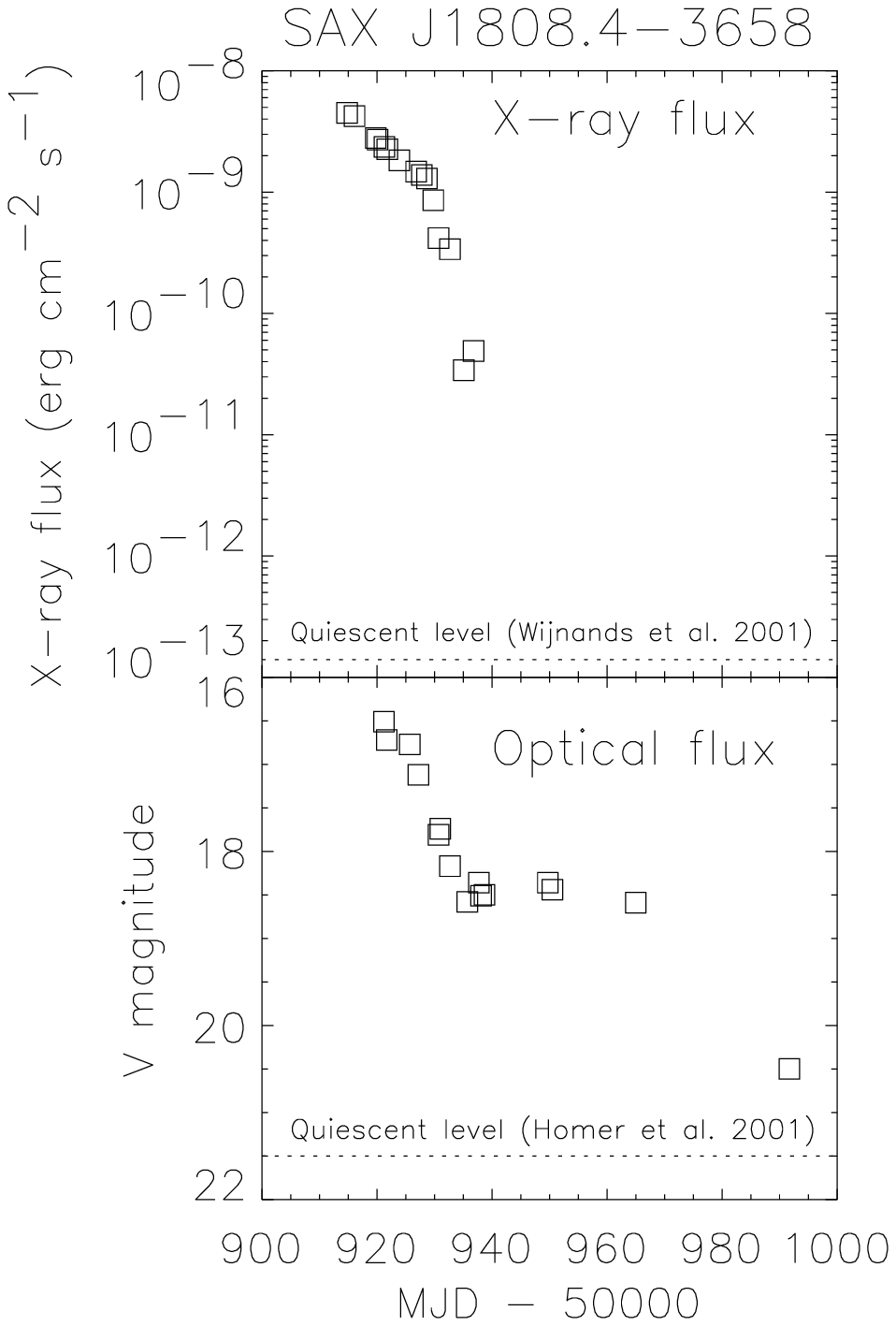}}
\figcaption{X-ray (3--150 keV) and optical flux histories during the 1998
outburst of SAX J1808.4$-$3658.  The quiescent levels well after the
outburst are also indicated.  The X-ray history (taken from Gilfanov
et al. 1998) contains a sharp break at MJD 50929, while the optical
history shows a break at MJD 50936.}
\vspace*{0.2in}

\noindent
of $10^8$ cm.  The outer disk will be cut off sharply near
the neutron star's tidal radius, $\approx R_{\rm Roche}$ (Frank et
al. 1992), which will in turn depend upon the mass ratio and thus
the binary inclination (Eggleton 1983).  We may infer the X-ray
luminosity from the X-ray flux through $L_{\rm x}=4\pi D^2 F_{\rm x}$,
and hence deduce the mass transfer rate $\dot M$.  Finally,
for the X-ray albedo of the disk, we use the results of previous
studies of X-ray reprocessing in LMXBs which found that $\eta_{\rm
d}\gtrsim 0.90$, indicating that only a small fraction of the incident
X-ray flux is absorbed by the accretion disk and reprocessed into the
optical band (Kallman, Raymond, \& Vrtilek 1991; de Jong, van
Paradijs, \& Augusteijn 1996). 

With the model parameters set as described above, we fit the heated
disk model to the observed photometry with two free parameters, $\cos 
i$ and the optical $V$-band extinction $A_V$.  
We computed our model fits on a grid with 99 values of $\cos i$ in the 
range 0.01--0.99 and 501 values of $A_V$ in the range 0.00--5.00.  
We computed the extinction in the other bands using the interstellar
reddening law of Rieke \& Lebofsky (1985).  For each $(A_V, \cos i)$
grid point, we fit the data for 10 different values of $\eta_{\rm d}$
in the range 0.90--0.99 and used only the best-fit value for that grid
point.  In order to ensure that we were working in the regime where
X-ray heating is important (and the X-ray flux is well determined), we
confined our fitting to the optical data prior to the break in the
X-ray light curve at MJD 50929.  \hfill The 

\centerline{\epsfxsize=8.5cm\epsfbox{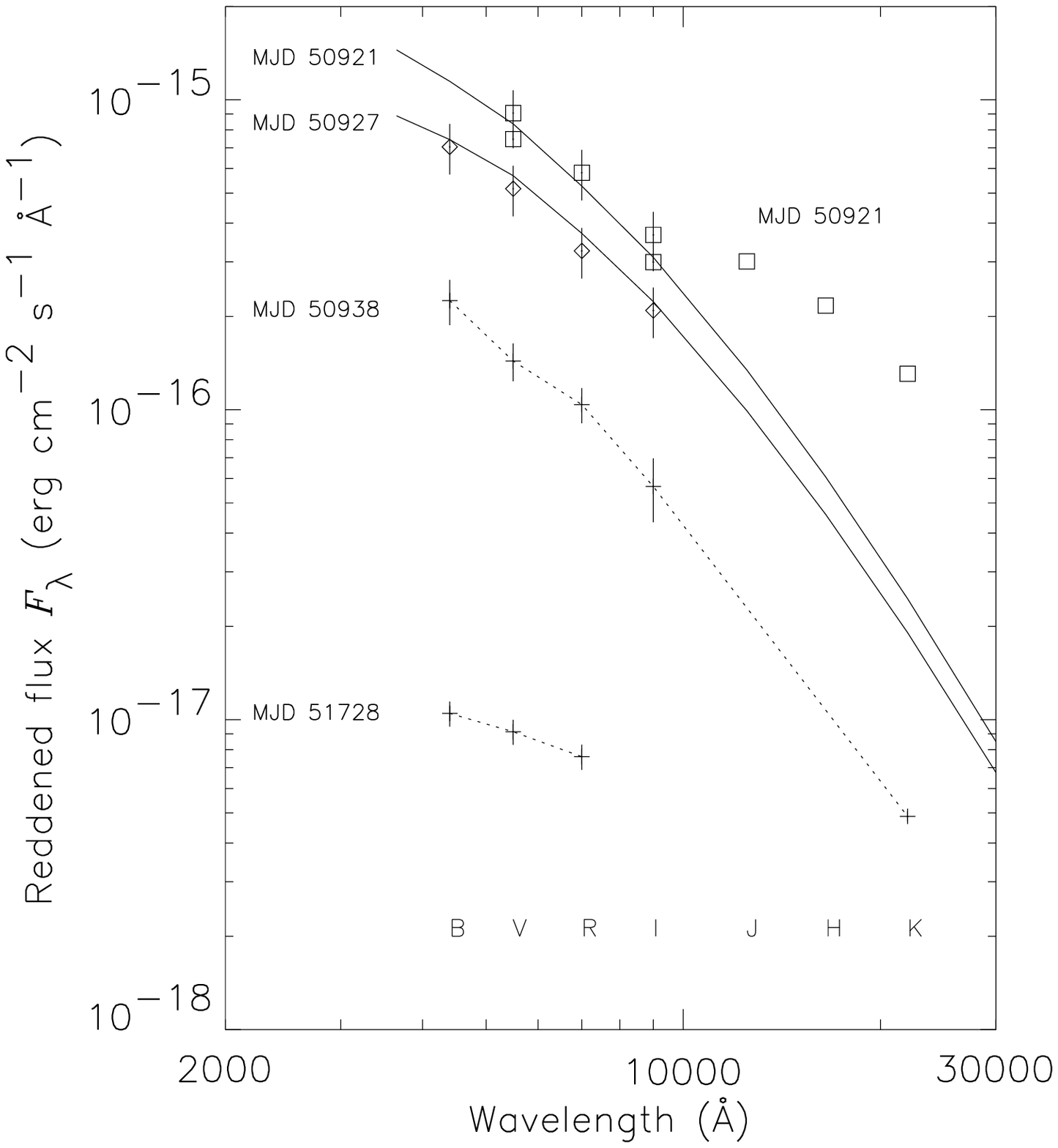}}
\figcaption{Broadband optical/IR spectra of SAX J1808.4$-3658$ at various
epochs during the 1998 outburst.  The solid curves are model fits,
while the dotted curves simply indicate a rough interpolation of the
data.  There is a clear IR excess with respect to an accretion disk
model on MJD 50921, but not on MJD 50938. } 

\vspace*{0.2in}
\noindent
fitting was performed simultaneously to all the {\em BVRI} data in
Table~1 prior to that date.

Our simple disk model was able to provide a good simultaneous
solution to these data.  The best-fit parameters were
$A_V=0.68^{+0.37}_{-0.28}$ and $\cos i= 0.65^{+0.23}_{-0.38}$, with
reduced $\chi^2=0.81$ (9 degrees of freedom), where the uncertainties
are quoted at the 90\%-confidence level.  The spectral model for two
epochs is shown by the solid curves in Figure~2, and a contour plot of
the allowed parameter space is shown in Figure~3.  The confidence
levels indicated by the contours in Figure~3 are determined as
described by Lampton, Margon, \& Bowyer (1976).  The hashed regions
in Figure~3 reflect the additional lower limit of $A_V>0.53$ set by
the measured Galactic dust extinction through the Galactic disk along
the line of sight (Schlegel, Finkbeiner, \& Davis 1998), and the
additional lower limit of $\cos i>0.15$ set by the absence of a deep
X-ray eclipse of the source (Chakrabarty \& Morgan 1998).  (We note
that the system must lie outside the Galactic disk, given its distance
and Galactic latitude.)  Accounting for these additional independent
limits, the best-fit parameter values are $A_V=0.68^{+0.37}_{-0.15}$
and $\cos i= 0.65^{+0.23}_{-0.33}$.  The allowed parameter space is
relatively narrow in $A_V$ with a central value only slightly larger
than the Galactic value.

To investigate how sensitive our conclusions are to the adopted source
distance $D=2.5$ kpc, we refit the data for distances of 2~kpc and
3~kpc as well.  For the larger distance, the confidence contours in
Figure~3 were slightly displaced upward and to the right, with
best-fit parameter values $A_V=0.74^{+0.39}_{-0.34}$ and $\cos
i=0.80^{+0.14}_{-0.38}$ with reduced $\chi^2=0.80$.  For 
the smaller distance, the contours were more elongated and displaced
downward and to the left, with best-fit parameter values
$A_V=0.63^{+0.31}_{-0.22}$ and $\cos i=0.39\pm 0.35$ with reduced
\hfill $\chi^2=0.85$.  

\centerline{\epsfxsize=8.5cm\epsfbox{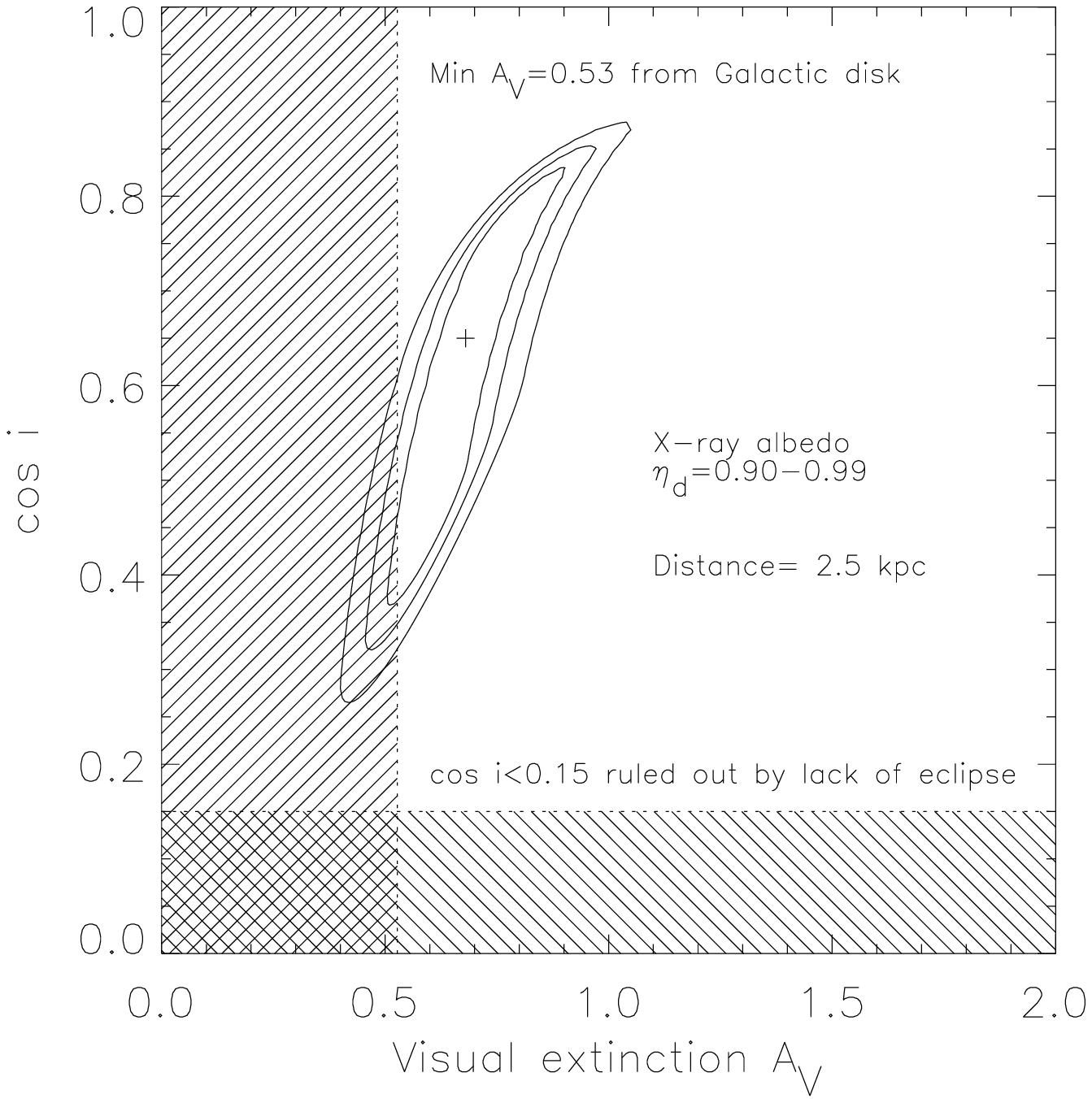}}
\figcaption{Confidence regions for $\cos i$ and $A_V$ from spectral
fitting.  The best-fit ($\cos i=0.65$, $A_V=0.68$) is indicated by the
cross, while the solid contours denote $\Delta\chi^2=$1.0, 2.3, 4.6
(corresponding to confidence levels of 39\%, 68\%, and 90\%).  The
hashed regions indicate addition limits from the line-of-sight
extinction in the Galactic disk and the lack of an X-ray eclipse.}
\vspace*{0.2in}

\noindent
In both cases, the fit values are quoted without
accounting for the addition independent limits discussed above.  The
assumed distance clearly plays a significant role in determining the
fit parameters, although the qualitative results of an extinction
value slightly greater than Galactic and an intermediate inclination
are robust.

\section{DISCUSSION}

We have shown that the optical spectrum of SAX J1808.4$-$3658 during
its 1998 outburst is well fit by an X-ray--heated accretion disk model.
The derived inclination range for the binary requires that the
companion mass is 0.05--0.10~$M_\odot$ (Chakrabarty \& Morgan 1998).
This is consistent with the low companion mass deduced from the long-term
average $\dot M$ for gravitational-radiation--driven mass transfer
(Chakrabarty \& Morgan 1998) and thus supports the recent prediction
that the mass donor is a low-mass brown dwarf (Bildsten \& Chakrabarty
2001).  This inclination range is also consistent with the
orbital-phase flux variability observed from the source in both the
X-ray (Chakrabarty \& Morgan 1998; Lee, Psaltis, \& Chakrabarty 2001,
in preparation) and optical (Giles et al. 1999; Homer et al. 2001)
bands. 

The strong infrared excess measured on MJD 50921 is clearly
inconsistent with emission from the X-ray heated disk (see Figure~2).
It is also orders of magnitude too bright to be due to the companion;
even X-ray heating does not mitigate this, due to the small solid
angle subtended by the star.  On the other hand, the infrared emission
on MJD 50938 is consistent with disk emission, indicating that the
cause of the earlier IR excess is transient in nature.  It is
interesting to note that the flux density of the IR excess is
comparable to the radio flux density measured from the source a week
later (Gaensler, Stappers, \& Getts 1999).  Radio/IR emission due to
synchrotron processes have been previously detected from some X-ray
binaries during outburst (see, e.g., Fender 2001).   The possibility
of a synchrotron origin for the IR excess on MJD 50921 (as well as the
radio emission) will be explored in detail elsewhere (Chakrabarty,
Gaensler, \& Stappers 2001, in preparation).   

\acknowledgements{We thank Phil Blanco, William Heindl, and Fred
Hammann of UCSD for obtaining and sharing their Keck observation with
us, and Sonja Vrielmann for obtaining the SAAO observation during outburst.
We also thank Rob Fender, Bryan Gaensler, Dimitrios Psaltis, Krzysztof
Stanek, and Ben Stappers for useful discussions.  This work was
supported in part by NASA under grant NAG5-9184 and contract
NAS8-38249.}

\end{document}